\RequirePackage[l2tabu, orthodox]{nag}
\RequirePackage{snapshot}

\documentclass[10pt,onecolumn]{extarticle}

\makeatletter
\if@twocolumn
  \usepackage[dvips,letterpaper,top=0.5in, bottom=0.5in, left=0.75in, right=0.5in,includefoot,heightrounded]{geometry}
\else
  \usepackage[dvips,letterpaper,margin=1in,includefoot,heightrounded]{geometry}
\fi

\usepackage{bm}

\usepackage{srcltx}

\usepackage{amsmath}
\usepackage{amssymb,amsfonts}

\usepackage{abstract}

\usepackage{graphicx}
\usepackage{epsfig}
\usepackage[usenames,dvipsnames]{color}
\usepackage[usenames,dvipsnames]{xcolor}
\usepackage{subfigure}

\usepackage{booktabs}

\usepackage{setspace}
\usepackage{flushend}
\usepackage{multicol}

\usepackage{cite}
\usepackage{url}
\usepackage[normalem]{ulem}

\usepackage{enumerate}

\usepackage{multirow}
\usepackage{array}

\usepackage{hyperref}

\usepackage{lipsum}

\usepackage[sc,tiny]{titlesec}

\usepackage{microtype}



\newcommand{\printtitle}{%
\makeatletter
\if@twocolumn

\twocolumn[%
  \maketitle
  \begin{onecolabstract}
    \myabstract
  \end{onecolabstract}
  \begin{center}
    \small
    \textbf{Keywords}
    \\\medskip
    \mykeywords
  \end{center}
  \bigskip
]
\saythanks
\else
  \maketitle
  \begin{abstract}
    \myabstract
  \end{abstract}
  \begin{center}
    \small
    \textbf{Keywords}
    \\\medskip
    \mykeywords
  \end{center}
  \bigskip
  \onehalfspacing
\fi
}

\title{%
A Multiplierless Pruned DCT-like Transformation for Image and Video Compression that Requires 10~Additions Only
}

\author{%
Vitor~A.~Coutinho
\thanks{
Vitor~A.~Coutinho
and
Renato~J.~Cintra
are
with the
Signal Processing Group,
Departamento de Estat\'istica
and
the Graduate Program in Electrical Engineering,
Universidade Federal de Pernambuco,
Recife, PE, Brazil
(e-mail: rjdsc@stat.ufpe.org).}
\and
Renato~J.~Cintra%
\and
F\'abio~M.~Bayer%
\thanks{%
F\'abio~M.~Bayer
is with the
Departamento de Estat\'istica
and Laborat\'orio de Ci\^encias Espaciais de Santa Maria (LACESM),
Universidade Federal de Santa Maria,
Santa Maria, RS, Brazil
(e-mail: bayer@ufsm.br).}
\and
Sunera~Kulasekera
\and
Arjuna~Madanayake%
\thanks{%
Sunera Kulasekera
and
Arjuna Madanayake
are with the
Department of Electrical and Computer Engineering,
The University of Akron, Akron, OH, USA
(e-mail: arjuna@uakron.edu).}
}

\date{}

\newcommand{\myabstract}{%
A multiplierless pruned approximate 8-point
discrete cosine transform~(DCT)
requiring only 10~additions
is introduced.
The proposed algorithm was
assessed
in
image and video compression,
showing competitive performance
with state-of-the-art methods.
Digital synthesis in
45\,nm~CMOS technology
up to place-and-route level indicates clock speed of 288 MHz at a 1.1~V supply.
The $8\times 8$ block rate is 36 MHz.
The DCT approximation
was  embedded into~HEVC reference software;
resulting video frames, at up to 327~Hz for 8-bit RGB HEVC,
presented
negligible
image degradation.
}

\newcommand{\mykeywords}{%
Approximate discrete cosine transform,
pruning,
pruned DCT,
HEVC
}

\begin{document}

\printtitle

\section{Introduction}
\label{intro}

The discrete cosine transform (DCT)
plays a fundamental role
in  signal processing techniques~\cite{rao1990discrete}
and is part of
modern
image and video standards,
such as
JPEG~\cite{Wallace1992}, %
MPEG-1~\cite{roma2007hybrid},
MPEG-2~\cite{mpeg2},
H.261~\cite{h261},
H.263~\cite{h263},
H.264/AVC~\cite{wiegand2003,h264},
and
the high efficiency video coding (HEVC)~\cite{hevc,hevc1}.
In particular,
the transform coding stage of the
H.264 and HEVC standards employs
the
8-point DCT of type~II~\cite{
hevc1,
sullivan2012}
among other transforms of different blocklenghts,
such as 4, 16, and 32~points~\cite{Bayer2013, Park2012, Park2013}.
In~\cite{Potluri2013},
the 8-point DCT stage of the HEVC was optimized.
Among the above-mentioned standards,
the HEVC
is capable of achieving high compression performance
at
approximately half the bit rate required by
H.264/AVC with same image quality~\cite{hevc1,Park2012,Ohm2012,Potluri2013}.
However,
HEVC possesses a significant computational
complexity in terms of arithmetic operations~\cite{Park2012, Ohm2012, sullivan2012, Potluri2013}.
In fact,
HEVC can be 2--4 times more computationally demanding
when compared to H.264/AVC~\cite{Park2012,Potluri2013}.
Therefore,
the proposal of efficient low-complexity
DCT-like approximations can
benefit future video codecs including emerging
HEVC-based systems.

Recently,
low-complexity DCT approximations have been considered
for image and video processing~\cite{haweel2001new, bas2008, bas2009, bas2013, cb2011, bc2012, Potluri2013, Bayer2013, Cintra2014-sigpro,tablada2015class}.
Such approximate transforms can offer
meaningful DCT estimations
at the expense of small errors.
Such trade-off is often acceptable
leading to low-power,
high-speed hardware realizations~\cite{Potluri2013},
while ensuring
adequate numerical accuracy.

In some applications,
such as data compression~\cite{Rao2001},
high-frequency components are often
zeroed by the quantization process.
Thus,
one may judiciously
restrict
the computation
to the quantities
that are likely
to be remain significant~\cite{Huang2000}.
This approach is called \emph{pruning}
and was originally proposed as a method
for computing the
discrete Fourier transform~(DFT)~\cite{Markel1971,Skinner1976}.

\subsection{Related Works}

In that context,
pruning was applied in time-domain,
i.e.,
particular input samples were ignored
and
the operations involving them were avoided~\cite{alves2000general}.
Frequency-domain pruning---discarding transform coefficients---is
an alternative approach.
This latter approach has been recently applied
in
mixed-radix FFT algorithms~\cite{wang2012generic},
cognitive radio design~\cite{airoldi2010energy},
and
wireless communications~\cite{whatmough2012vlsi}.
Another example of a pruning-like algorithm is
the well-known Goertzel method
for DFT computation~\cite{Oppenheim2010,kim2011islanding,carugati2012variable}.

For the DCT case,
pruning
was originally proposed by Wang in~\cite{wang1991pruning}
considering a decimation-in-time algorithm
for power-of-two blocklengths.
In~\cite{skodras1994fast},
such algorithm was generalized for arbitrary
blocklength.
In~\cite{Lecuire2012},
Lecuire~\emph{et al.}
extended the pruning
method
to the
two-dimensional case,
referring to the method as the \emph{zonal} DCT,
which is an alternative terminology.
In~\cite{Karakonstantis2009},
Karakonstantis~\emph{et al.}
proposed
a hardware-based pruning approach for the DCT computation.
Instead of discarding high frequency DCT coefficients,
a VLSI system capable
of computing
low frequency components using faster paths
was suggested.
Such method was applied
in the context of voltage scaling for low-power dissipation.
In the context
of low-powered wireless vision sensor networks,
a pruned approximate DCT was proposed
in~\cite{kouadria2013low}
based on
the DCT approximation theory advanced in~\cite{bas2013}.
In~\cite{meher2014efficient}
the pruning terminology was
employed in a different context.
It was considered to describe
a hardware computation of the DCT
which maintains
the system word size constant
by means of a
controlled discarding of least-significant bits.

\subsection{Aims}

In response to the growing need
for high compression ratios for image and moving pictures
as prescribed in~\cite{hevc},
we propose
a further reduction
of the computational cost
of the DCT computation
in the context of JPEG- and HEVC-like
coding and processing.
The goal of this paper
is to offer a comprehensive
analysis
of pruning schemes
in combination with approximate transforms.
The sough schemes
must be
capable of
reducing
the number of computed approximate DCT coefficients
and,
at the same time,
the effected degradation
on picture quality
must be negligible.

In the present work,
a multiplication-free pruned approximate 8-point DCT
is sought.
We
also
aim at
VLSI realizations
of both \mbox{1-D} and \mbox{2-D}
versions of the proposed pruned approximate transform.
The sought methods
are intended to be fully embedded into
an open source HEVC reference software~\cite{hevcreference}
for
performance assessment
in real time video coding.

\section{Proposed pruned approximate DCT}

\subsection{Proposed approximation}

In~\cite{bc2012}
a very low-complexity 8-point
DCT approximation was introduced
and it is
referred to as the modified rounded DCT (RDCT),
which is associated to the following low-complexity matrix:
\begin{align}
\bm{T} =
\left[
\begin{array}{rrrrrrrr}
1 & \phantom{-}1 & 1 & 1 & 1 & \phantom{-}1 & \phantom{-}1 & 1 \\
1 & 0 & 0 & 0 & 0 & 0 & 0 &-1 \\
1 & 0 & 0 &-1 &-1 & 0 & 0 & 1 \\
0 & 0 &-1 & 0 & 0 & 1 & 0 & 0 \\
1 &-1 &-1 & 1 & 1 &-1 &-1 & 1 \\
0 &-1 & 0 & 0 & 0 & 0 & 1 & 0 \\
0 &-1 & 1 & 0 & 0 & 1 &-1 & 0 \\
0 & 0 & 0 &-1 & 1 & 0 & 0 & 0
\end{array}
\right]
.
\end{align}
Its associated
fast algorithm requires only 14~additions,
having the lowest computational complexity among
the meaningful DCT approximations archived
in literature~\cite{Britanak2007,haweel2001new,bas2008,bas2009,bas2013,cb2011,
bc2012, Potluri2013}.
Considering the orthogonalization methods
described in~\cite{cintra2011integer},
an orthonormal approximation for the DCT is given by
given by:

\begin{align}
\hat{\bm{C}}
=
\bm{D}\cdot\bm{T}
=
\frac{1}{2}
\cdot
\operatorname{diag}
\left(
\frac{1}{\sqrt{2}}, \sqrt{2}, 1, \sqrt{2},
\frac{1}{\sqrt{2}}, \sqrt{2}, 1, \sqrt{2},
\right)
\cdot\bm{T} ,
\end{align}
where
$\operatorname{diag}(\cdot)$
returns a diagonal matrix with the elements of its argument.

\begin{figure}[t]
\centering
\input{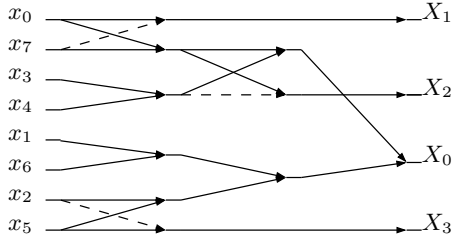}
\caption{Signal flow graph
relating
input data $x_n$, $n=0,1,\ldots,7$,
to output $X_k$, $k=0,1,3,4$,
according to
$\bm{T}_4$.
Dashed arrows are multiplications by $-1$}
\label{figure-fast-algorithm}
\end{figure}

By means of
analyzing
fifty
512$\times$512 8-bit representative
standard images~\cite{uscsipi},
we noticed
that
the
\mbox{2-D} version of the
8-point modified RDCT~\cite{bc2012}
can concentrate
in average $\approx$98\% of the total image energy
in the 16~lower frequency coefficients.
Additionally,
in JPEG-like image compression,
the quantization step
is prone to zero the high frequency
coefficients~\cite{Lecuire2012}.
Therefore,
computational efforts may be saved
by not computing the high frequency coefficients,
keeping only low-frequency coefficients.
These considerations
yield
the following transformation
derived from
the low-complexity matrix
associated to
the modified RDCT:
\begin{align}
\bm{T}_4 =
\left[
\begin{array}{rrrrrrrr}
1 & \phantom{-}1 & 1 & 1 & 1 & \phantom{-}1 & \phantom{-}1 & 1 \\
1 & 0 & 0 & 0 & 0 & 0 & 0 &-1 \\
1 & 0 & 0 &-1 &-1 & 0 & 0 & 1 \\
0 & 0 &-1 & 0 & 0 & 1 & 0 & 0 \\
\end{array}
\right]
.
\end{align}
Above transformation computes
the four lower frequency components
of the \mbox{1-D}
original
modified RDCT
low-complexity matrix,
which corresponds to the 16~lower frequency components
of the associated \mbox{2-D} version.
Thus,
considering the orthogonalization methods
described in~\cite{cintra2011integer},
we can obtain a semi-orthogonal matrix
given by:
\begin{align}
\hat{\bm{C}}_4
=
\bm{D}_4 \cdot\bm{T}_4
=
\frac{1}{2}
\cdot
\operatorname{diag}
\left(
\frac{1}{\sqrt{2}}, \sqrt{2}, 1, \sqrt{2}
\right)
\cdot\bm{T}_4 .
\end{align}
Matrix $\hat{\bm{C}}_4$
is the pruned version of the modified RDCT.
For image and video compression,
the scaling diagonal matrix~$\bm{D}_4$
does not introduce any computational overhead,
since
it can be merged
into the quantization step~\cite{bas2008,bas2009,cb2011,bc2012,Potluri2013}.
Therefore,
in such context,
the computational complexity of~$\hat{\bm{C}}_4$
is essentially confined into
the low-complexity matrix~$\bm{T}_4$.
Aiming at the efficient of implementation
of~$\bm{T}_4$,
we factorized it in a product of
extremely low-complexity sparse matrices.
Such factorization
is based on decimation-in-time methods
as described in~\cite{Blahut2010,Britanak2007,cb2011}.
Thus,
the following expression is obtained:
\begin{align}
\bm{T}_4
=
& \bm{P} \cdot \bm{A}_3 \cdot \bm{A}_2 \cdot \bm{A}_1
,
\end{align}
where
\begin{align}
\begin{split}
\bm{P}  &=
\begin{bmatrix}
     1  &  0   & 0 &   0\\
     0  &  0  &  0  &  1\\
     0  &  1  &  0  &  0\\
     0  &  0  &  1  &  0
\end{bmatrix} ,
\\
\bm{A}_3  &=
\begin{bmatrix}
     1  &  1  &  0  &  0  &  0 \\
     0  &  0  &  1  &  0  &  0 \\
     0  &  0 &   0 &   1  &  0 \\
     0 &   0 &   0  &  0  &  1
\end{bmatrix},
\\
\bm{A}_2  &=
\begin{bmatrix}
        1  &  \phantom{-}0  &  \phantom{-}0  &  \phantom{-}1  &  \phantom{-}0  &  \phantom{-}0 \\
        0  &  \phantom{-}1  &  \phantom{-}1  &  \phantom{-}0  &  \phantom{-}0  &  \phantom{-}0 \\
        1  &  \phantom{-}0  &  \phantom{-}0  & -1  &  \phantom{-}0  &  \phantom{-}0 \\
       0  &  \phantom{-}0 &   \phantom{-}0  &  \phantom{-}0  & -1   & \phantom{-}0 \\
       0  &  \phantom{-}0  &  \phantom{-}0  &  \phantom{-}0  &  \phantom{-}0  &  \phantom{-}1
\end{bmatrix},
\\
\bm{A}_1 &=
\begin{bmatrix}
        1  &  \phantom{-}0  &  \phantom{-}0   & \phantom{-}0  &  \phantom{-}0  &  \phantom{-}0  &  \phantom{-}0  &  \phantom{-}1\\
        0  &  \phantom{-}1  &  \phantom{-}0  &  \phantom{-}0  &  \phantom{-}0  &  \phantom{-}0  &  \phantom{-}1  &  \phantom{-}0\\
        0  &  \phantom{-}0  &  \phantom{-}1  &  \phantom{-}0  &  \phantom{-}0  &  \phantom{-}1  &  \phantom{-}0  &  \phantom{-}0\\
        0  &  \phantom{-}0  &  \phantom{-}0  &  \phantom{-}1  &  \phantom{-}1  &  \phantom{-}0  &  \phantom{-}0  &  \phantom{-}0\\
        0  & \phantom{-}0  &  \phantom{-}1  &  \phantom{-}0  &  \phantom{-}0 &  -1  &  \phantom{-}0  &  \phantom{-}0\\
        1  &  \phantom{-}0  &  \phantom{-}0  &  \phantom{-}0  &  \phantom{-}0  &  \phantom{-}0  & \phantom{-} 0  & -1
\end{bmatrix},
\end{split}
\end{align}
where
$\bm{A}_1$,
$\bm{A}_2$,
and
$\bm{A}_3$,
are
pre-addition matrices~\cite{Blahut2010}
and
$\bm{P}$ is a permutation matrix.
Fig.~\ref{figure-fast-algorithm}
provides
the signal flow graph of the fast algorithm for~$\bm{T}_4$,
relating input signal $x_n$, $n=0,1,\ldots,7$
to output signal $X_k$, $k=0,1,\ldots,7$.
Transform-domain components $X_4,\ldots,X_7$ are
not represented,
being set to zero.

Based on the \mbox{2-D} computation of the DCT~\cite{Britanak2007},
the approximate \mbox{2-D} DCT is given by~\cite{bas2008}:
\begin{align}
\bm{B}
=
\hat{\bm{C}}
\cdot
\bm{A}
\cdot
\hat{\bm{C}}^\top
,
\end{align}
where $\bm{A}$ is
an input 8$\times$8 image subblock,
$\bm{B}$ is the associated
transform-domain 8$\times$8 output image subblock
and superscript $\top$ indicates matrix transposition.
For instance,
JPEG-like schemes are entirely based
on the DCT-based transformation of 8$\times$8 subblocks~\cite{Wallace1992}.

In a similar fashion,
the \mbox{2-D} forward pruned
transformation can be derived~\cite{Makkaoui2010,Lecuire2012,kouadria2013low}
and
it is
described according to:
\begin{equation}
\bm{B}'
=
\hat{\bm{C}}_4
\cdot
\bm{A}
\cdot
\hat{\bm{C}}_4^\top
,
\end{equation}
where
$\bm{B}'$ is the
pruned
4$\times$4  output image subblock.

Matrix~$\bm{B}'$ contains
a subset of the transform-domain
coefficients~$\bm{B}$
furnished by the modified RDCT.
The approximate \mbox{2-D} spectrum
is given by
the 8$\times$8 matrix~$\hat{\bm{B}}$
constituted of~$\bm{B}'$ in upper-left corner
and zeros elsewhere,
as shown below:
\begin{align}
\bm{B}
\approx
\hat{\bm{B}}
=
\left[
\begin{array}{c|c}
\bm{B}' & \bm{0}_4 \\
\hline
\bm{0}_4 & \bm{0}_4
\end{array}
\right]
,
\end{align}
where
$\bm{0}_4$
represents the 4$\times$4 null matrix.
The inverse transformation
can be computed by taking the inverse transformation
of the above zero-padded matrix~$\hat{\bm{B}}$.
However,
this is equivalent to the following computation:
\begin{equation}
\bm{A}
\approx
\hat{\bm{A}}
=
\hat{\bm{C}}_4^\top
\cdot
\bm{B}'
\cdot
\hat{\bm{C}}_4
.
\end{equation}
Therefore,
padding becomes unnecessary.
Additionally,
we noticed that
$\hat{\bm{C}}_4^\top$
is the Moore-Penrose pseudo-inverse of
$\hat{\bm{C}}_4$~\cite{seber2008matrix}.

\subsection{Complexity Assessment}

By counting
the number of
multiplications,
additions,
and
bit-shifting operations,
we assessed the computational cost
of the proposed \mbox{1-D} and \mbox{2-D} pruned modified RDCT.
\tablename~\ref{table-complexity} compares
the obtained complexities with
the computational costs
associated to
traditional
and
state-of-the-art
DCT methods.
Selected DCT approximations
include:
(i)~the signed DCT (SDCT)~\cite{haweel2001new};
(ii)~the rounded DCT (RDCT)~\cite{cb2011};
(iii)~the modified RDCT~\cite{bc2012};
and
a set of DCT approximations introduced by Bouguezel-Ahmad-Swamy,
namely,
BAS-2008~\cite{bas2008},
BAS-2009~\cite{bas2009},
and
BAS-2013~\cite{bas2013}.
Here,
we also included the computational cost of the \emph{exact}
DCT computation
according to
Chen's DCT algorithm~\cite{chen1977},
which is the algorithm employed in the
HEVC
codec~\cite{hevcreference}.
Each of the selected methods
was assessed
both in its full and pruned versions.
The \mbox{1-D} and \mbox{2-D} versions
retained the four
and 16 lower frequency coefficients, respectively.

\begin{table*}
\centering
\caption{Computational complexity assessment}
\label{table-complexity}
\begin{tabular}{l  ccc | ccc}
\toprule
 &
\multicolumn{3}{c}{Nonpruned} &
\multicolumn{3}{c}{Pruned}
\\
\cmidrule {2-7}
1-D Method & Mult. & Add. & Shift & Mult. & Add. & Shift
 \\
\midrule
DCT (by definition)  &
64 & 56 & 0 & 32 & 28 & 0
\\
Chen's DCT~\cite{chen1977} &
16 & 26 & 0 & 6 & 12 & 0 \\
SDCT~\cite{haweel2001new} &
0 & 24 & 0 & 0 & 20 & 0 \\
BAS-2008~\cite{bas2008} &
0 & 18 & 2 & 0 & 14 & 1 \\
BAS-2009~\cite{bas2009} &
0 & 18 & 0 & 0 & 14 & 0 \\
BAS-2013~\cite{bas2013,kouadria2013low} &
0 & 24 & 0 & 0 & 20 & 0\\
RDCT~\cite{cb2011} &
0 & 22 & 0 & 0 & 16 & 0 \\
Modified RDCT~\cite{bc2012} &
0 & 14 & 0 & 0 & \textbf{10} & 0 \\
\midrule
2-D Method & Mult. & Add. & Shift & Mult. & Add. & Shift
 \\
\midrule
DCT (by definition)  &
1024 & 896 & 0 & 384 & 336 & 0
\\
Chen's DCT~\cite{chen1977} &
256 & 416 & 0 & 72 & 144 & 0 \\
SDCT~\cite{haweel2001new} &
0 & 384 & 0 & 0 & 240 & 0 \\
BAS-2008~\cite{bas2008} &
0 & 288 & 32 & 0 & 168 & 12 \\
BAS-2009~\cite{bas2009} &
0 & 288 & 0 & 0 & 168 & 0 \\
BAS-2013~\cite{bas2013,kouadria2013low} &
0 & 384 & 0 & 0 & 240 & 0 \\
RDCT~\cite{cb2011} &
0 & 352 & 0 & 0 & 192 & 0 \\
Modified RDCT~\cite{bc2012} &
0 & 224 & 0 & 0 & \textbf{120} & 0\\
\bottomrule
\end{tabular}
\end{table*}

The proposed \mbox{1-D}
method demands
\emph{only 10~additions}.
The associate percent complexity reduction compared to selected state-of-art
methods is presented in \tablename~\ref{table-complexity_reduction},
for both the \mbox{1-D} and \mbox{2-D} case.
A certain
arithmetic complexity reduction
was already expected by using pruning approach.
However,
when considering the \mbox{2-D} transformation,
arithmetic complexity
reduction effected by the pruning procedure
is even more significant,
as shown in \tablename~\ref{table-complexity_reduction}.
In fact,
the 8$\times$8 nonpruned
\mbox{2-D} transformation can be decomposed into eight
nonpruned row-wise \mbox{1-D} transformations;
followed by eight column-wise instantiations of
the same \mbox{1-D} transformation.
In contrast,
the
proposed pruned \mbox{2-D} transformation
can be decomposed into eight pruned row-wise
\mbox{1-D} transformations of the rows;
followed by \emph{only four} pruned \mbox{1-D} transformations.
Therefore,
the pruned \mbox{2-D} transformation
calls the \mbox{1-D} algorithm fewer times
when compared with the nonpruned case.
The complexity values presented in
Table~\ref{table-complexity}
were calculated according to
above considerations.
As a consequence,
the proposed transformation requires \emph{120 additions}.
The proposed method outperforms
the recently proposed pruned approximation
described in~\cite{kouadria2013low},
requiring 50\% less operations
for both \mbox{1-D} and \mbox{2-D} versions.
Moreover,
the comparison among pruned-only versions of above methods
shows
that
the proposed approximation
demands
28{.}5\%
less operations,
in both \mbox{1-D} and \mbox{2-D} cases,
than
the best competing methods,
namely BAS-2008~\cite{bas2008} and BAS-2009~\cite{bas2009}.

\begin{table}
\centering
\caption{Percent complexity reduction of the proposed method compared to state-of-art methods}
\begin{tabular}{l c c}
\toprule
Method & 1-D & 2-D \\
\midrule
SDCT~\cite{haweel2001new} & 58{.}3\% & 68{.}8\% \\
BAS-2008~\cite{bas2008} & 50{.}0\% & 62{.}5\% \\
BAS-2009~\cite{bas2009} & 44{.}4\% & 58{.}3\% \\
BAS-2013~\cite{bas2013} & 58{.}3\% & 68{.}8\% \\
RDCT~\cite{cb2011} & 54{.}5\% & 65{.}9\% \\
Modified RDCT~\cite{bc2012} & 28{.}6\% & 46{.}4\% \\
\bottomrule
\end{tabular}
\label{table-complexity_reduction}
\end{table}

\section{Image compression}

We
processed
the set of images
mentioned in the previous section
according to
the
image
compression simulation
detailed in~\cite{bas2008,cb2011,bc2012}.
Images were subdivided
into 8$\times$8 blocks and were submitted
to \mbox{2-D} transformation
according to
the proposed pruned approximate DCT
and competing methods.
The resulting coefficients in the transform domain
were submitted
to the standard quantization operation
for luminance~\cite[p.~155]{bhaskaran1997}.
We adopted a variable length coding approach,
where the number of zeroed transform-domain coefficients
is determined by the quantization step.
The maximum number of non-zero coefficients
is~16,
as imposed by the pruning scheme.
Subsequently,
inverse transformations were considered.
For the proposed method,
the inverse procedure
described in Section~II
was applied
and
compressed images were reconstructed.
Original and processed images
were evaluated for image degradation
using
the peak signal-to-noise ratio~(PSNR)~\cite[p.~9]{bhaskaran1997} and Structural Similarity~(SSIM)~\cite{Wang2004}.
We also computed the number of zeros (NZ)
after quantization,
which provides the percentage number of zeroed coefficients after quantization step
and furnishes a measure of energy compaction
in the transform domain.
High values of NZ translates into longer runs of zeros,
which are beneficial for subsequent run-length encoding
and Huffman coding stages~\cite{bhaskaran1997}	.

\begin{table*}
\centering
\caption{Performance assessment in image compression}
\label{table-performance}
\begin{tabular}{l   ccc |  ccc}
\toprule
\multirow{3}{*}{Method} &
\multicolumn{3}{c}{Nonpruned} &
\multicolumn{3}{c}{Pruned}
\\
\cmidrule{2-4} \cmidrule{5-7}
 & PSNR & SSIM & NZ (\%) & PSNR &SSIM& NZ (\%)\\
\midrule
Chen's DCT~\cite{chen1977} &
33.10 & 0.90 & 81.83 &
30.40 & 0.86 & 86.19 \\
SDCT~\cite{haweel2001new} &
29.28 & 0.84 & 80.20 &
27.14 & 0.77 & 86.27 \\
BAS-2008~\cite{bas2008} &
32.17 & 0.89 & 80.87 &
29.24 & 0.83 & 86.00 \\
BAS-2009~\cite{bas2009} &
31.72 & 0.88 & 80.59 &
28.69 & 0.82 & 86.16\\
BAS-2013~\cite{bas2013,kouadria2013low} &
31.82 & 0.88 & 80.52 &
28.72 & 0.82 & 86.10\\
RDCT~\cite{cb2011} &
31.91 & 0.88 & 81.03 &
28.93 & 0.82 & 86.45\\
Modified RDCT~\cite{bc2012} &
30.94 & 0.86 & 79.83 &
26.37 & 0.72 & \textbf{86.75}\\
\bottomrule
\end{tabular}
\end{table*}

\begin{figure}[h]
\centering
\subfigure[Modified RDCT~\cite{bc2012} (nonpruned)]{\epsfig{file=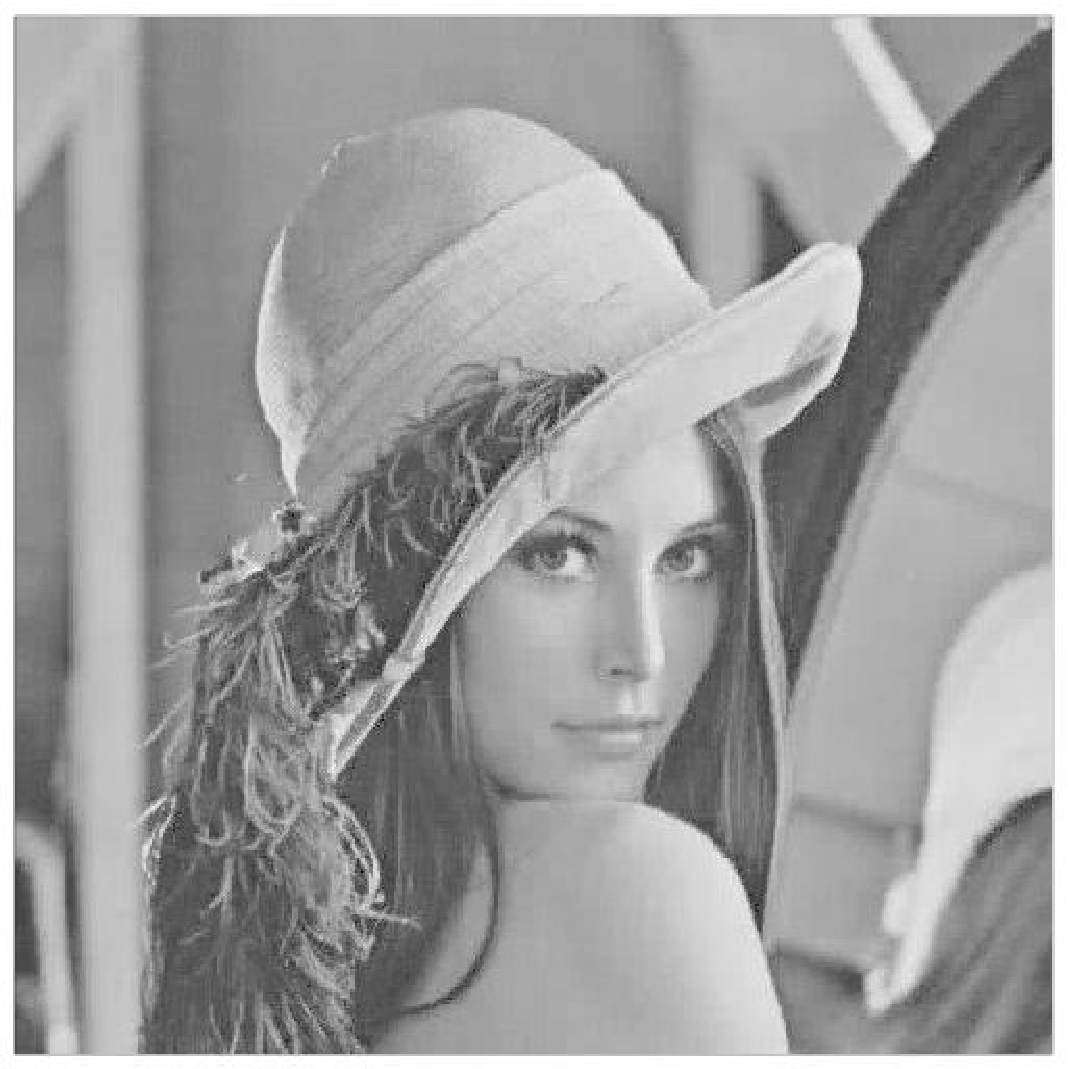,width=0.28\linewidth}}
\subfigure[Proposed pruned transform]{\epsfig{file=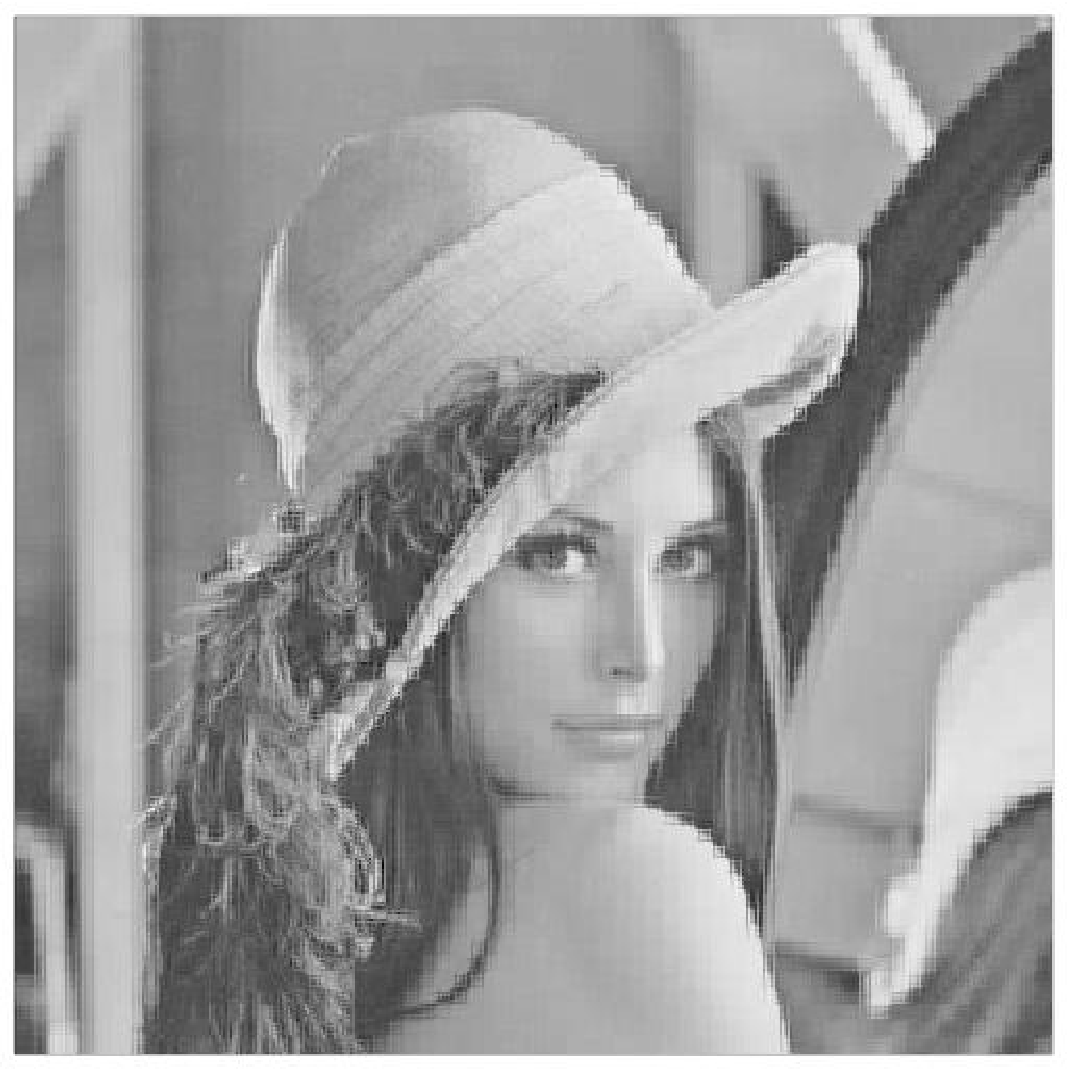,width=0.28\linewidth}}
\caption{Compressed Lena images.}
\label{figure-lena}
\end{figure}

In contrast with~\cite{bas2008, bas2009,bas2013,kouadria2013low},
we adopted average measurements,
which are less prone to variance effects and
fortuitous data.
Table~\ref{table-performance}
shows the average PSNR values
and percent values for NZ
based on the selected image set
for each considered method.
Results indicate that
the proposed method can significantly
reduce the computational complexity,
while maintaining good PSNR figures.
For instance,
considering the original and pruned MRDCT,
we noticed a $\approx 15\%$ decrease in PSNR and SSIM;
however the associated arithmetic complexity reduction is of $\approx 50\%$.

A qualitative comparison between the Lena~\cite{uscsipi} compressed image
obtained from the above describe procedure
using the modified RDCT~\cite{bc2012} and the proposed pruned transform
is shown in Fig.~\ref{figure-lena}.

\section{VLSI architectures}

To further investigate the capabilities of the proposed algorithm,
we separate the modified RDCT and
the proposed pruned approximation
for hardware synthesis
and
evaluation in the actual HEVC scheme.

\begin{table*}
\centering
\caption{Resource consumption on Xilinx XC6VLX240T-1FFG1156 de\-vice.}
\label{fpga}
\begin{tabular}{lcccccc}
\toprule
Method &
CLB &
FF &
CPD &
$F_\text{max}$\! &
$D_p$\! &
$Q_p$\!
\\
\midrule
Modified RDCT & 445 & 1696 & 3.390 & 294.98 & 2.74 & 3.44 \\
\hline
Proposed      & 247  & 961 & 2.946 & 339.44 & 1.35 & 3.43 \\
\bottomrule
\end{tabular}
\end{table*}

\begin{table*}
\centering
\caption{Resource consumption for CMOS 45\,nm synthesis.}
\label{asic}

\begin{tabular}
{l
>{\centering}m{13pt}
>{\centering}m{13pt}
>{\centering}m{15pt}
>{\centering}m{9pt}
>{\centering}m{15pt}
>{\centering}m{15pt}
m{15pt}<{}}
\toprule
Method &
 Area &
AT &
$\mathrm{AT}^2$ &
\strut\!\!CPD &
$F_\text{max}$ &
$D_p$ &
$Q_p$\!
\\
\midrule
Modified RDCT & 0.073  & 0.261 & 0.936 & 3.582 & 279.17 & 0.050 & 0.039 \\
\hline
Proposed      & 0.043  & 0.149 & 0.518 & 3.471 & 288.10 & 0.012 & 0.011  \\
\bottomrule
\end{tabular}
\end{table*}

\subsection{FPGA Architecture}

These approximations were realized as
a separable \mbox{2-D} block transform using
two \mbox{1-D} transform blocks and a transpose buffer.
Such blocks were initially modeled and tested
in Matlab Simulink and
then
combined
to furnish the complete \mbox{2-D} transform.
The resulting architecture
was physically realized on
a Xilinx Virtex-6 XC6VLX240T-1FFG1156
field programmable gate array (FPGA)
device and
validated using
hardware-in-the-loop testing through the JTAG interface.
The
DCT approximation
FPGA prototype was verified using more than 10000~test vectors
with complete agreement with theoretical values.
Quantities were obtained
from the Xilinx FPGA synthesis by
accessing the \texttt{xflow.results} report file for each run of the
design flow.
Metrics,
including
configurable logic blocks (CLB) and flip-flop (FF)
count,
critical path delay (CPD)~(in~ns),
and
maximum operating frequency ($F_\text{max}$, in~MHz),
are provided.
In addition,
static ($Q_p$, in~W) and
frequency normalized
dynamic power ($D_p$, in~mW/MHz) consumptions
were
estimated using the Xilinx XPower Analyzer.
\subsection{CMOS Place-Route}

Following FPGA based verification,
the hardware description language code was ported to 45\,nm~CMOS
technology and subject to synthesis and place-and-route steps using Cadence Encounter.
Both FPGA synthesize and CMOS place-and-route results are tabulated
in Table~\ref{fpga} and~\ref{asic},
respectively.
For the CMOS place-and-route,
critical path delay (CPD)~(in~ns),
area (in~$\mathrm{mm}^2$),
area-time complexity (AT, in~$\mathrm{mm}^2\cdot\mathrm{ns}$),
area-time-squared complexity (AT$^2$,
in $\mathrm{mm}^2\cdot\mathrm{ns}^2$),
maximum operating frequency ($F_\text{max}$, in~MHz),
static ($Q_p$, in~W) and
frequency normalized
dynamic power ($D_p$, in~mW/MHz) consumptions
are also provided.
The FPGA realization of the proposed
pruned DCT approximation
showed
a reduction of 44.49\% in area as measured
by the number of CLBs
and a
50.72\% reduction in
frequency normalized dynamic power consumption
when
compared with the full DCT approximation.
Synthesis at the 45\,nm CMOS technology node using FreePDK45
standard cells revealed a
41.09\% reduction in area
and
a 76\% reduction in
frequency normalized dynamic power
for a supply voltage fixed at
$V_\text{DD}=1.1\,\mathrm{V}$.
All metrics indicate clear advantages of using
the proposed pruned DCT approximation over
the full 8-point approximation.
Further,
the 288~MHz~CMOS clock indicates
a block rate of 36~MHz
and
a frame-rate of 327~Hz,
assuming
8-bit RGB video at
1920$\times$1080 resolution.

\section{HEVC software simulation}

\begin{figure}[hh]
\subfigure[]{\epsfig{file=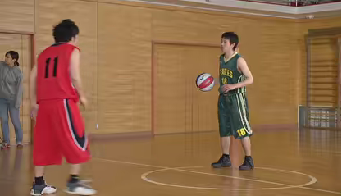,width=0.45\linewidth}}
\subfigure[]{\epsfig{file=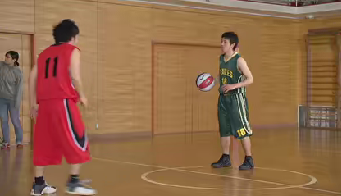,width=0.45\linewidth}}
\centering
\subfigure[]{\epsfig{file=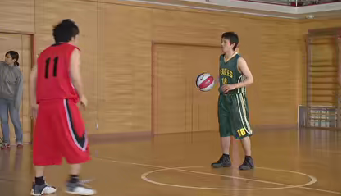,width=0.45\linewidth}}
\caption{Selected frame from `BasketballPass' test video
coded by means of
(a)~the Chen's DCT algorithm~(PSNR~37{.}62~dB),
(b)~the modified RDCT~(PSNR~37{.}42~dB),
and
(c)~the proposed pruned approximation, with 76{.}2\% less arithmetic operations then Chen's DCT~(PSNR 37{.}41~dB).}
\label{figure-frame}
\end{figure}

We considered
real time video coding
by embedding the proposed algorithms
into
the  HEVC reference software
by the
Fraunhofer Heinrich Hertz Institute~\cite{hevcreference}.
The original transform included in the HEVC reference software
is a scaled approximation of Chen's DCT algorithm.
Our methodology
consists of replacing
the 8$\times$8 DCT algorithm of the reference software
by the modified RDCT and the proposed pruned approximation.

\begin{figure}[hh]
\subfigure[PSNR values related to bit rate]{\epsfig{file=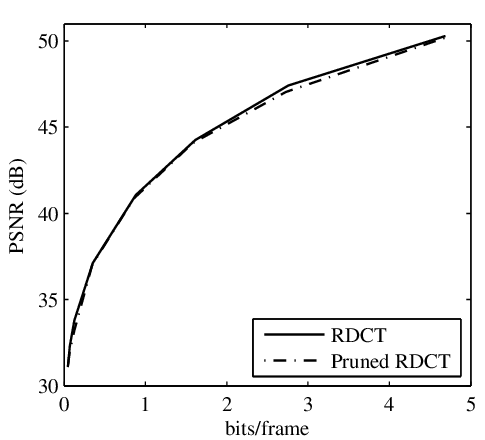,width=0.45\linewidth}\label{figure-rd-curve}}
\subfigure[PSNR values related to QP]{\epsfig{file=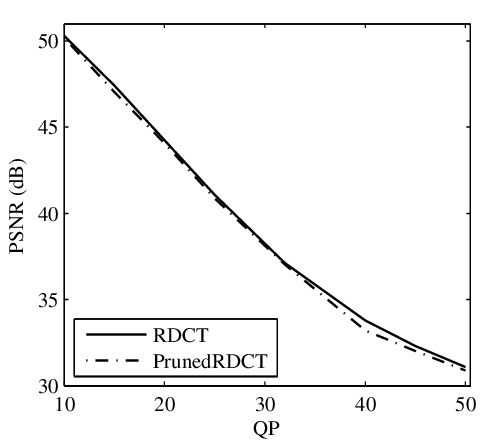,width=0.45\linewidth}\label{figure-rd-curve-qp}}
\caption{RD curves for `BasketballPass' test sequence.}
\label{figure-rd}
\end{figure}

Fig.~\ref{figure-frame} shows three 416$\times$240 frames
of the `BasketballPass' test sequence~\cite{hevseq}
obtained from the HEVC simulation.
Resulting frames were coded using
the Chen's DCT algorithm (Fig.~\ref{figure-frame}(a)),
the modified RDCT (Fig.~\ref{figure-frame}(b)),
and the proposed pruned approximation (Fig.~\ref{figure-frame}(c)).
The PSNR values for these three frames are shown in Fig.~\ref{figure-frame}.
The pruned approximation
effected minimal image degradation---less than 0.25\,dB.
On the other hand, computational complexity of
the 8-point DCT was
significantly reduced---76{.}2\% less arithmetic operations
when compared with
the original Chen's DCT algorithm.

We have also computed rate distortion~(RD) curves
for both RDCT and the proposed pruned approximation
using standard video sequences~\cite{hevseq}.
For such,
we varied
the quantization point~(QP)
from 0 to 50
and
computed
the PSNR of the proposed pruned
approximate with reference to the RDCT along with the bits/frame of the encoded video.
As a result,
we obtained the curves shown in
Fig.~\ref{figure-rd-curve}.
The PSNR values related to QP are shown in~\figurename~\ref{figure-rd-curve-qp}.
The difference in the rate points
between
the RDCT and the proposed pruned approximation is
less than 0.57dB, which is
smaller than $1.3~\%$.

\section{Conclusion}

In this paper,
we presented a
\emph{very} low-complexity DCT approximation
obtained via pruning.
The resulting approximate transform requires
\emph{only}~10~additions
and
possesses performance metrics comparable with state-of-the-art methods,
including the recent architecture presented in~\cite{kouadria2013low}.
The proposed pruning approach can be adapted to
other transformation methods,
regardless of the transform size.
By means of
computational simulation,
VLSI hardware realizations,
and
a full
HEVC
implementation,
we demonstrated the practical relevance of our method
as an image and video codec.
Our goal with the image and video simulations is \emph{not}
to suggest the modification of existing standards,
which would be unfeasible.
Instead
we aim at showing that
(i)~pruned approximations can be considered
in tailored low-complexity, low-power systems
for accelerated decoding of JPEG and HEVC
and
(ii)~approximation methods combined with pruning
are a viable alternative to the design
of future standards.
For future work,
we intend to apply the pruned approach to other
discrete transforms methods for
different blocklengths.
In particular,
the 4-, 16-, and 32-point DCT-based approximations
are naturally fitted to the proposed approach.
Moreover,
a prospective
study on the energy distribution
in the transform domain
could indicate the optimal number of coefficients to
retain in the pruning process.
Forthcoming applications include
low-power wireless vision sensor networks
and
accelerated image decoding.

\section*{Acknowledgments}
Authors acknowledge partial support
from
CNPq,
FACEPE,
FAPERGS,
and
The University of Akron.

{\small
\bibliographystyle{IEEEtran}
\bibliography{dct-clean}

\begin{thebibliography}{10}
\providecommand{\url}[1]{#1}
\csname url@samestyle\endcsname
\providecommand{\newblock}{\relax}
\providecommand{\bibinfo}[2]{#2}
\providecommand{\BIBentrySTDinterwordspacing}{\spaceskip=0pt\relax}
\providecommand{\BIBentryALTinterwordstretchfactor}{4}
\providecommand{\BIBentryALTinterwordspacing}{\spaceskip=\fontdimen2\font plus
\BIBentryALTinterwordstretchfactor\fontdimen3\font minus
  \fontdimen4\font\relax}
\providecommand{\BIBforeignlanguage}[2]{{%
\expandafter\ifx\csname l@#1\endcsname\relax
\typeout{** WARNING: IEEEtran.bst: No hyphenation pattern has been}%
\typeout{** loaded for the language `#1'. Using the pattern for}%
\typeout{** the default language instead.}%
\else
\language=\csname l@#1\endcsname
\fi
#2}}
\providecommand{\BIBdecl}{\relax}
\BIBdecl

\bibitem{rao1990discrete}
K.~R. Rao and P.~Yip, \emph{Discrete Cosine Transform: Algorithms, Advantages,
  Applications}.\hskip 1em plus 0.5em minus 0.4em\relax San Diego, CA: Academic
  Press, 1990.

\bibitem{Wallace1992}
G.~K. Wallace, ``The {JPEG} still picture compression standard,'' \emph{IEEE
  Transactions on Consumer Electronics}, vol.~38, pp. xviii--xxxiv, 1992.

\bibitem{roma2007hybrid}
N.~Roma and L.~Sousa, ``Efficient hybrid {DCT}-domain algorithm for video
  spatial downscaling,'' \emph{EURASIP Journal on Advances in Signal
  Processing}, vol. 2007, pp. 30--30, 2007.

\bibitem{mpeg2}
{International Organisation for Standardisation}, ``Generic coding of moving
  pictures and associated audio information -- {P}art 2: Video,'' ISO, {ISO/IEC
  JTC1/SC29/WG11} - Coding of Moving Pictures and Audio, 1994.

\bibitem{h261}
{International Telecommunication Union}, ``{ITU}-{T} recommendation {H}.261
  version 1: Video codec for audiovisual services at $p \times 64$ kbits,''
  ITU-T, Tech. Rep., 1990.

\bibitem{h263}
------, ``{ITU}-{T} recommendation {H}.263 version 1: Video coding for low bit
  rate communication,'' ITU-T, Tech. Rep., 1995.

\bibitem{wiegand2003}
T.~Wiegand, G.~J. Sullivan, G.~Bjontegaard, and A.~Luthra, ``Overview of the
  {H}.264/{AVC} video coding standard,'' \emph{IEEE Transactions on Circuits
  and Systems for Video Technology}, vol.~13, pp. 560--576, 2003.

\bibitem{h264}
J.~V. Team, ``Recommendation {H}.264 and {ISO}/{IEC} 14 496--10 {AVC}: Draft
  {ITU}-{T} recommendation and final draft international standard of joint
  video specification,'' ITU-T, Tech. Rep., 2003.

\bibitem{hevc}
{International Telecommunication Union}, ``High efficiency video coding:
  Recommendation {ITU-T H.265},'' ITU-T Series H: Audiovisual and Multimedia
  Systems, Tech. Rep., 2013.

\bibitem{hevc1}
M.~T. Pourazad, C.~Doutre, M.~Azimi, and P.~Nasiopoulos, ``{HEVC}: The new gold
  standard for video compression: How does {HEVC} compare with {H.264/AVC}?''
  \emph{IEEE Consumer Electronics Magazine}, vol.~1, pp. 36--46, Jul. 2012.

\bibitem{sullivan2012}
G.~J. Sullivan, J.~Ohm, W.~Han, and T.~Wiegand, ``Overview of the high
  efficiency video coding ({HEVC}) standard,'' \emph{IEEE Transactions on
  Circuits and Systems for Video Technology}, vol.~22, pp. 1649--1668, Dec.
  2012.

\bibitem{Bayer2013}
F.~M. Bayer, R.~J. Cintra, A.~Madanayake, and U.~S. Potluri, ``Multiplierless
  approximate 4-point {DCT} {VLSI} architectures for transform block coding,''
  \emph{Electronics Letters}, vol.~49, pp. 1532--1534, 2013.

\bibitem{Park2012}
J.~Park, W.~Nam, S.~Han, and S.~Lee, ``2-{D} large inverse transform
  (16$\times$16, 32$\times$32) for {HEVC} ({H}igh {E}fficiency {V}ideo
  {C}oding),'' \emph{Journal of Semiconductor Technology and Science}, vol.~2,
  pp. 203--211, 2012.

\bibitem{Park2013}
S.~Park and P.~K. Meher, ``Flexible integer {DCT} architectures for {HEVC},''
  in \emph{IEEE International Symposium on Circuits and Systems (ISCAS)}, 2013,
  pp. 1376--1379.

\bibitem{Potluri2013}
U.~S. Potluri, A.~Madanayake, R.~J. Cintra, F.~M. Bayer, S.~Kulasekera, and
  A.~Edirisuriya, ``Improved 8-point approximate {DCT} for image and video
  compression requiring only 14 additions,'' \emph{IEEE Transactions on
  Circuits and Systems I: Regular Papers}, vol.~61, no.~6, pp. 1727--1740,
  2014.

\bibitem{Ohm2012}
J.~Ohm, G.~J. Sullivan, H.~Schwarz, T.~K. Tan, and T.~Wiegand, ``Comparison of
  the coding efficiency of video coding standards - including {H}igh
  {E}fficiency {V}ideo {C}oding ({HEVC}),'' \emph{IEEE Transactions on Circuits
  and Systems for Video Technology}, vol.~22, pp. 1669--1684, Dec. 2012.

\bibitem{haweel2001new}
T.~I. Haweel, ``A new square wave transform based on the {DCT},'' \emph{Signal
  Processing}, vol.~82, pp. 2309--2319, 2001.

\bibitem{bas2008}
S.~Bouguezel, M.~O. Ahmad, and M.~N.~S. Swamy, ``Low-complexity 8$\times$8
  transform for image compression,'' \emph{Electronics Letters}, vol.~44, pp.
  1249--1250, Sep. 2008.

\bibitem{bas2009}
------, ``A fast 8$\times$8 transform for image compression,'' in
  \emph{International Conference on Microelectronics (ICM)}, Dec. 2009, pp.
  74--77.

\bibitem{bas2013}
------, ``Binary discrete cosine and {H}artley transforms,'' \emph{IEEE
  Transactions on Circuits and Systems I: Regular Papers}, vol.~60, pp.
  989--1002, 2013.

\bibitem{cb2011}
R.~J. Cintra and F.~M. Bayer, ``A {DCT} approximation for image compression,''
  \emph{IEEE Signal Processing Letters}, vol.~18, pp. 579--582, Oct. 2011.

\bibitem{bc2012}
F.~M. Bayer and R.~J. Cintra, ``{DCT}-like transform for image compression
  requires 14 additions only,'' \emph{Electronics Letters}, vol.~48, pp.
  919--921, 2012.

\bibitem{Cintra2014-sigpro}
R.~J. Cintra, F.~M. Bayer, and C.~J. Tablada, ``Low-complexity 8-point {DCT}
  approximations based on integer functions,'' \emph{Signal Processing},
  vol.~99, pp. 201--214, 2014.

\bibitem{tablada2015class}
\BIBentryALTinterwordspacing
C.~J. Tablada, F.~M. Bayer, and R.~J. Cintra, ``A class of {DCT} approximations
  based on the {F}eig-{W}inograd algorithm,'' \emph{Signal Processing}, vol.
  113, pp. 38--51, 2015. [Online]. Available:
  \url{http://www.sciencedirect.com/science/article/pii/S0165168415000341}
\BIBentrySTDinterwordspacing

\bibitem{Rao2001}
K.~R. Rao and P.~Yip, \emph{The Transform and Data Compression Handbook}.\hskip
  1em plus 0.5em minus 0.4em\relax {CRC} Press {LLC}, 2001.

\bibitem{Huang2000}
Y.~Huang, J.~Wu, and C.~Chang, ``A generalized output pruning algorithm for
  matrix-vector multiplication and its application to compute pruning discrete
  cosine transform,'' \emph{IEEE Transactions on Signal Processing}, vol.~48,
  pp. 561--563, 2000.

\bibitem{Markel1971}
J.~Markel, ``{FFT} pruning,'' \emph{IEEE Transactions on Audio and
  Electroacoustics}, vol.~19, pp. 305--311, 1971.

\bibitem{Skinner1976}
D.~P. Skinner, ``Pruning the decimation in-time {FFT} algorithm,'' \emph{{IEEE}
  Transactions on Acoustics, Speech and Signal Processing}, vol.~24, pp.
  193--194, 1976.

\bibitem{alves2000general}
R.~G. Alves, P.~L. Osorio, and M.~N.~S. Swamy, ``General {FFT} pruning
  algorithm,'' in \emph{43rd IEEE Midwest Symposium on Circuits and Systems},
  vol.~3, 2000, pp. 1192--1195.

\bibitem{wang2012generic}
L.~Wang, X.~Zhou, G.~E. Sobelman, and R.~Liu, ``Generic mixed-radix {FFT}
  pruning,'' \emph{IEEE Signal Processing Letters}, vol.~19, pp. 167--170, Mar.
  2012.

\bibitem{airoldi2010energy}
R.~Airoldi, O.~Anjum, F.~Garzia, A.~M. Wyglinski, and J.~Nurmi,
  ``Energy-efficient fast {Fourier} transforms for cognitive radio systems,''
  \emph{IEEE Micro}, vol.~30, pp. 66--76, Nov. 2010.

\bibitem{whatmough2012vlsi}
P.~N. Whatmough, M.~R. Perrett, S.~Isam, and I.~Darwazeh, ``{VLSI} architecture
  for a reconfigurable spectrally efficient {FDM} baseband transmitter,''
  \emph{IEEE Transactions on Circuits and Systems I: Regular Papers}, vol.~59,
  pp. 1107--1118, May 2012.

\bibitem{Oppenheim2010}
A.~Oppenheim and R.~Schafer, \emph{Discrete-Time Signal Processing},
  3rd~ed.\hskip 1em plus 0.5em minus 0.4em\relax Pearson, 2010.

\bibitem{kim2011islanding}
J.~H. Kim, J.~G. Kim, Y.~Ji, Y.~Jung, and C.~Won, ``An islanding detection
  method for a grid-connected system based on the {G}oertzel algorithm,''
  \emph{IEEE Transactions on Power Electronics}, vol.~26, pp. 1049--1055, Apr.
  2011.

\bibitem{carugati2012variable}
I.~Carugati, S.~Maestri, P.~G. Donato, D.~Carrica, and M.~Benedetti, ``Variable
  sampling period filter {PLL} for distorted three-phase systems,'' \emph{IEEE
  Transactions on Power Electronics}, vol.~27, pp. 321--330, Jan. 2012.

\bibitem{wang1991pruning}
Z.~Wang, ``Pruning the fast discrete cosine transform,'' \emph{IEEE
  Transactions on Communications}, vol.~39, pp. 640--643, May 1991.

\bibitem{skodras1994fast}
A.~N. Skodras, ``Fast discrete cosine transform pruning,'' \emph{IEEE
  Transactions on Signal Processing}, vol.~42, pp. 1833--1837, Jul. 1994.

\bibitem{Lecuire2012}
V.~Lecuire, L.~Makkaoui, and J.-M. Moureaux, ``Fast zonal {DCT} for energy
  conservation in wireless image sensor networks,'' \emph{Electronics Letters},
  vol.~48, pp. 125--127, 2012.

\bibitem{Karakonstantis2009}
G.~Karakonstantis, N.~Banerjee, and K.~Roy, ``Process-variation resilient and
  voltage-scalable {DCT} architecture for robust low-power computing,''
  \emph{IEEE Transactions on Very Large Scale Integration ({VLSI}) Systems},
  vol.~18, pp. 1461--1470, 2009.

\bibitem{kouadria2013low}
N.~Kouadria, N.~Doghmane, D.~Messadeg, and S.~Harize, ``Low complexity {DCT}
  for image compression in wireless visual sensor networks,'' \emph{Electronics
  Letters}, vol.~49, pp. 1531--1532, 2013.

\bibitem{meher2014efficient}
P.~K. Meher, S.~Y. Park, B.~K. Mohanty, K.~S. Lim, and C.~Yeo, ``Efficient
  integer {DCT} architectures for {HEVC},'' \emph{IEEE Transactions on Circuits
  and Systems for Video Technology}, vol.~24, pp. 168--178, Jan. 2014.

\bibitem{hevcreference}
{Joint Collaborative Team on Video Coding (JCT-VC)}, ``{HEVC} reference
  software documentation,'' Fraunhofer Heinrich Hertz Institute, Tech. Rep.,
  2013.

\bibitem{Britanak2007}
V.~Britanak, P.~Yip, and K.~R. Rao, \emph{Discrete Cosine and Sine Transforms:
  General Properties, Fast Algorithms and Integer Approximation}.\hskip 1em
  plus 0.5em minus 0.4em\relax Elsevier, 2007.

\bibitem{cintra2011integer}
\BIBentryALTinterwordspacing
R.~J. Cintra, ``An integer approximation method for discrete sinusoidal
  transforms,'' \emph{Journal of Circuits, Systems, and Signal Processing},
  vol.~30, pp. 1481--1501, 2011. [Online]. Available:
  \url{http://www.springerlink.com/content/nw5u0267254t3683/}
\BIBentrySTDinterwordspacing

\bibitem{uscsipi}
``The {USC}-{SIPI} image database,'' \url{http://sipi.usc.edu/database/}, 2011,
  {U}niversity of {S}outhern {C}alifornia, {S}ignal and {I}mage {P}rocessing
  {I}nstitute.

\bibitem{Blahut2010}
R.~E. Blahut, \emph{Fast Algorithms for Signal Processing}.\hskip 1em plus
  0.5em minus 0.4em\relax Cambridge University Press, 2010.

\bibitem{Makkaoui2010}
L.~Makkaoui, V.~Lecuire, and J.~Moureaux, ``Fast zonal {DCT}-based image
  compression for wireless camera sensor networks,'' \emph{2nd International
  Conference on Image Processing Theory Tools and Applications ({IPTA})}, pp.
  126--129, 2010.

\bibitem{seber2008matrix}
G.~A.~F. Seber, \emph{A Matrix Handbook for Statisticians}.\hskip 1em plus
  0.5em minus 0.4em\relax John Wiley \& Sons, Inc, 2008.

\bibitem{chen1977}
W.~H. Chen, C.~Smith, and S.~Fralick, ``A fast computational algorithm for the
  {D}iscrete {C}osine {T}ransform,'' \emph{IEEE Transactions on
  Communications}, vol.~25, no.~9, pp. 1004--1009, Sep. 1977.

\bibitem{bhaskaran1997}
V.~Bhaskaran and K.~Konstantinides, \emph{Image and Video Compression
  Standards}.\hskip 1em plus 0.5em minus 0.4em\relax Boston: Kluwer Academic
  Publishers, 1997.

\bibitem{Wang2004}
Z.~Wang, A.~C. Bovik, H.~R. Sheikh, and E.~P. Simoncelli, ``Image quality
  assessment: from error visibility to structural similarity,'' \emph{IEEE
  Transactions on Image Processing}, vol.~13, pp. 600--612, Apr. 2004.

\bibitem{hevseq}
``{HEVC} {T}est {V}ideo {S}equence,''
  ftp://hvc:US88Hula@ftp.tnt.uni-hannover.de/testsequences, 2013, heinrich
  Hertz Institute.

\end{thebibliography}
}

\end{document}